# *Analysis of Raman and Ellipsometric Responses of $Nb_xW_{1-x}Se_2$ alloys*

Albert F. Rigosi[1], Heather M. Hill[1], Sergiy Krylyuk[2], Nhan V. Nguyen[1], Angela R. Hight Walker[1], Albert V. Davydov[1], and David B. Newell[1]

[1]National Institute of Standards and Technology, Gaithersburg, MD 20899, USA

[2]Theiss Research, La Jolla, CA 92037, USA

david.newell@nist.gov

**Abstract:** The growth of transition metal dichalcogenide (TMDC) alloys provides an opportunity to experimentally access information elucidating how optical properties change with gradual substitutions in the lattice compared with their pure compositions. In this work, we performed growths of alloyed crystals with stoichiometric compositions between pure forms of $NbSe_2$ and $WSe_2$, followed by an optical analysis of those alloys by utilizing Raman spectroscopy and spectroscopic ellipsometry.

## 1. Introduction

The category of materials known as transition metal dichalcogenides (TMDCs) has been the subject of intense study in recent years, with latest efforts focusing on the quantum confinement in two dimensions resulting in direct bandgaps [1-7] and well-defined many-body effects [8-10]. Because TMDCs have wide-ranging applicability in photovoltaics, spintronics, optoelectronics, and energy storage [11-14], the ability to control and manipulate various facets of their inherent properties is strongly desired. Furthermore, this family of materials can be catalogued into several groups based on their characteristic electrical behavior. Semiconducting TMDCs, such as $WSe_2$, exhibit phenomena including coupled spin and valley degrees of freedom [15-18], whereas metallic TMDCs, such as $NbSe_2$, display properties such as charge density waves and superconductivity [19-27]. When alloyed, these materials offer advantages ranging from the control of a semiconductor's carrier density, construction of flexible gas sensors, and fabrication of 1D edge contacts or junctions [32-36]. Though the pure forms of $NbSe_2$ and $WSe_2$ have been well-studied optically [37-43], their alloys need more assessments [32-33, 44-47].

In this work, three optical spectroscopies were used to evaluate alloyed $Nb_xW_{1-x}Se_2$ crystals grown using chemical vapor transport (CVT). Samples exfoliated from these crystals were characterized with energy-dispersive X-ray spectroscopy (EDS) and Raman spectroscopy. An analysis of the optical, in-plane dielectric function $\varepsilon(E)$ was conducted using data acquired with spectroscopic ellipsometry (SE).

## 2. Experimental Methods

### 2.1 Crystal Growth and Energy-Dispersive X-ray Spectroscopy

$Nb_xW_{1-x}Se_2$ single-crystals on the centimeter scale were grown using CVT. Appropriate amounts of Nb (99.9 %), W (99.9 %) and Se (99.999 %) powders with a total weight of about 1 g were vacuum-sealed in a quartz ampoule with approximately 100 mg of $SeBr_4$ (99 %) as a transport agent. All precursors were purchased from Strem Chemicals, Inc.[1] Next, the ampoules were placed in a single-zone furnace so that the temperature at the charge zone and the growth zone was 980 °C and 850 °C, respectively. The ampoules were cooled down by switching off the furnace after 6 d of growth. All grown crystals were mechanically exfoliated onto fused silica substrates. Due to the limitations on both the equipment and the solid solubility, concentrations of Nb in the mid-range were unable to successfully be grown.

EDS analyses were performed on the bulk crystals directly after growth using a JEOL JSM-7100F field-emission scanning electron microscope equipped with an Oxford Instruments X-Max 80 silicon drift detector and a FEI Quanta 200 scanning electron microscope equipped with a Bruker XFlash 5030 silicon drift detector at a 10 keV accelerating voltage.

### 2.2 Raman Spectroscopy

Raman spectroscopy was performed on freshly exfoliated crystals (taken from bulk crystals) of each composition (approximately 100 nm thick). Raman spectra were collected with a Renishaw InVia micro-Raman spectrometer using a 514.5-nm wavelength excitation laser source. The laser was fixed at a constant power of 500 µW and had a spot size of about 1 µm. Acquisition times varied between 10 min and 30 min, and the optical path included a 50 × objective and 1200 $mm^{-1}$ grating. Rectangular Raman maps were collected at room temperature in a backscattering configuration with step sizes of 5 µm in a 5 by 3 raster-style grid.

### 2.3 Spectroscopic Ellipsometry

SE measurements were performed on freshly exfoliated crystals (taken from bulk crystals) all compositions with a Woollam M-2000 Ellipsometer, Model XI-210, and its 75 W Xenon light source comprising wavelengths between about 210 nm to 1000 nm (1.24 eV to 5.88 eV) with 1.25 nm resolution. The elliptical spot size measured approximately 100 µm along the semi-minor axis and 200 µm to 300 µm along the semi-major axis. SE measures the change in phase and polarization state of light reflected from a

---

sample of interest. Each crystal was measured over 20 times to directly extract $\varepsilon(E)$. Data were acquired at two angles of incidence (60° and 70°) and subsequently converted by the software from Fresnel reflection coefficients, representative of $p$ and $s$ polarized light ($R_p$ and $R_s$), to the related quantities psi and delta ($\Psi$ and $\Delta$), where $\frac{R_p}{R_s} = e^{i\Delta} \tan \Psi$. The dielectric function of each alloyed crystal was then converted from psi and delta directly.

## 3. Characterization of Alloyed Crystals

EDS data are shown in Fig. 1 on a logarithmic vertical scale and are taken for each of the $Nb_xW_{1-x}Se_2$ crystals of mixed composition, with the main relevant responses occurring between 1 keV and 2.5 keV. The response from Se (green band) remains unchanged as expected, whereas the W (purple band) decreases in relative intensity as the material approaches pure $NbSe_2$. In the case of Nb (orange band) the signal strength does decrease with Nb atom concentration, but for the top two panels, he presented concentrations of $x^{(1)}$ and $x^{(2)}$ are Nb peaks that were too difficult to distinguish since the detection limit of the measurement was exceeded. Though the Nb peaks still appear albeit just barely above the detection limit, determining an accurate concentration of Nb is not feasible from these data.

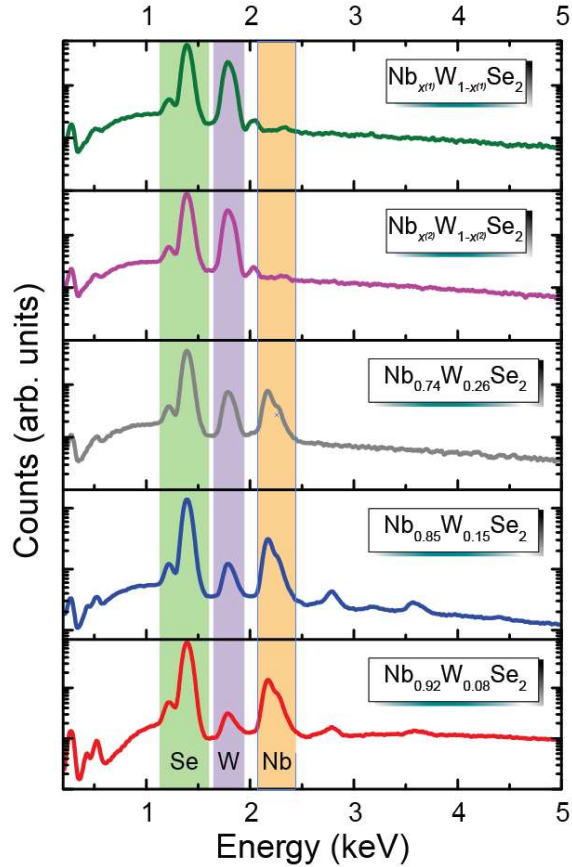

**Fig. 1.** EDS spectra are shown for the various CVT-grown crystals of mixed composition with the concentration labeled on the right (values of $x$ include $x^{(1)}$, $x^{(2)}$, 0.74, 0.85, and 0.92). The primary Se, W, and Nb responses are marked within the green, purple, and orange shaded regions, respectively. One can extract the compositions for the bottom three panels with better accuracy than the top two panels, whose compositions are not directly measurable.

As established, Raman spectroscopy is a noninvasive technique that can probe a variety of 2D materials on various substrates over large distances [48-50]. Figure 2 (a) shows the Raman spectra of the various modes which arise for the CVT-grown crystals with their corresponding Nb concentrations ($x$) labeled on the right. In all spectra, Raman modes were fit with Lorentzian peaks to extract the Raman shift and the full width at half maximum (FWHM). The pink shaded region highlights the most dynamic changes in the spectra. Each pure TMDC exhibits its $A_{1g}$ and $E_{2g}$ modes within this region, with $WSe_2$ also exhibiting the 2LA(M) mode at approximately 262 cm$^{-1}$. In the case of the $WSe_2$ $A_{1g}$ and $E_{2g}$ modes, located at approximately 258 cm$^{-1}$ and 249 cm$^{-1}$, respectively, they persist until Nb concentration values above $x = 0.74$ is achieved, with the latter mode persisting until W concentrations fall beneath 8 % ($x = 0.92$). A similar observation is made with the $A_{1g}$ and $E_{2g}$ modes of $NbSe_2$, with the former (at about 232 cm$^{-1}$) being the first to vanish as $x$ decreases and the latter (at about 241 cm$^{-1}$) persisting for low values of $x$ and finally vanishing in the case of pure $WSe_2$. These trends in Raman shift are shown in Fig. 2 (b). For Fig. 2 (c), the same five modes have their FWHM displayed as a function of $x$. Though many of the peak widths remain relatively stable in the interval $0 < x < 0.92$, some of the peaks demonstrate an abrupt increase in width as $x$ approaches 1.

Modes for WSe$_2$ at approximately 360 cm$^{-1}$, 375 cm$^{-1}$, and 395 cm$^{-1}$ are of second-order, and their respective assignments follow. The first mode is either a 2E$_{1g}$(Γ) or an A$_{1g}$(M)+TA(M) [48], with an alternate assignment of 2E'(Γ)–LA(K) and 2A$_1$'(Γ)–LA(K) [40]. The second mode has been reported as an E$_{2g}$(M)+LA(M), E'(Γ)+LA(M), and A$_1$'(Γ)+LA(M), and the third mode consists of an E'(Γ)+LA(K), A$_1$'(Γ)+LA(K), and 3LA(M) [40, 51]. These first two modes more rapidly vanish with increasing Nb concentration, but the third mode appears to persist up to about $x = 0.74$. Comparison of these modes with those in the literature is crucial to help verify that the alloys are taking on the approximate stoichiometric values for which they have been grown.

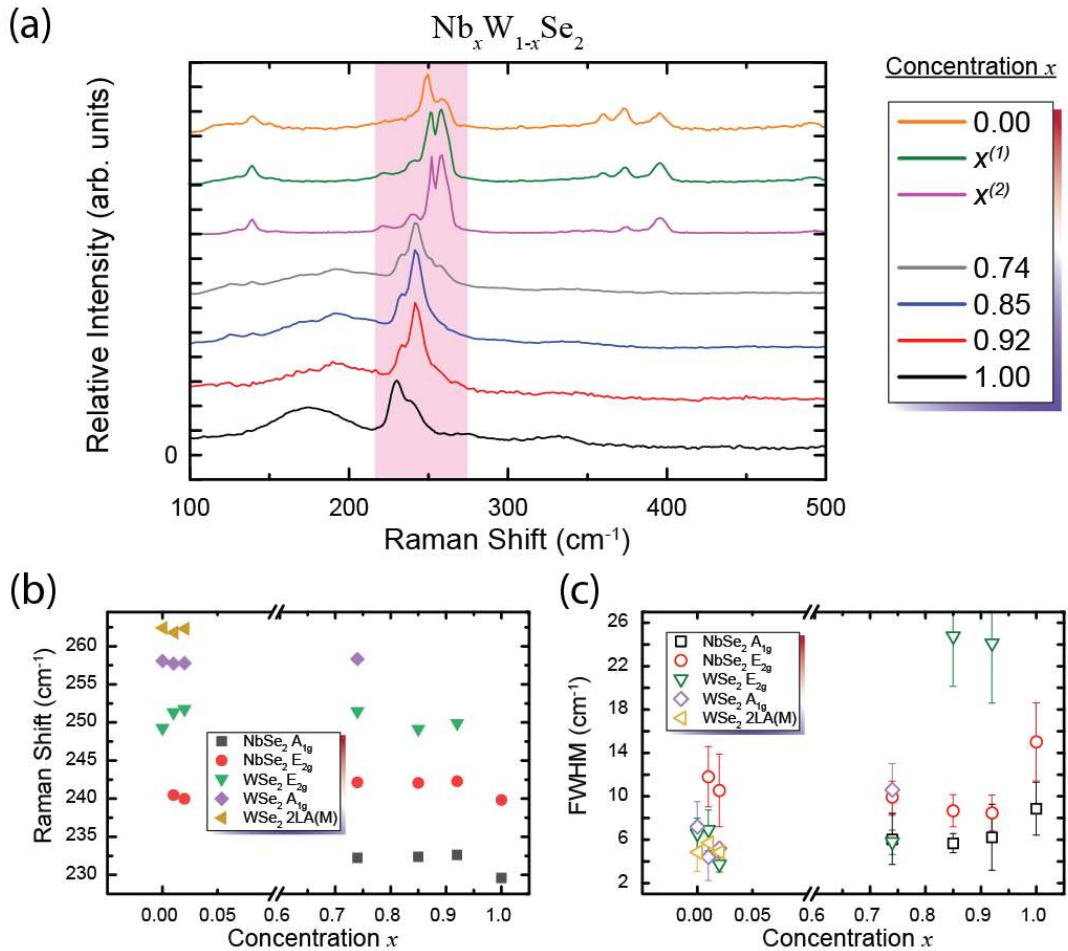

**Fig. 2.** (a) Raman spectra are shown for the various CVT-grown crystals with the concentration labeled on the right (values of $x$ include 0, $x^{(1)}$, $x^{(2)}$, 0.74, 0.85, 0.92, and 1). (b) The five modes of interest in the pink shaded region of (a) were fit with Lorentzian peaks to extract the shift and FWHM. A$_{1g}$ and E$_{2g}$ modes for both NbSe$_2$ and WSe$_2$ (and the 2LA(M) for the latter) are plotted as a function of Nb concentration $x$. (c) The FWHM of each mode is plotted as a function of $x$. The symbols (and their outer colors) are identical to their corresponding representations in (b). Error bars for the FWHM data represent the 1σ uncertainty of the width of the peak based on the Lorentzian fits.

## 4. Dielectric Properties

### 4.1 SE Measurements

By performing SE measurements, additional optical properties such as the dielectric function ($\varepsilon(E) = \varepsilon_1(E) + i\varepsilon_2(E)$), refractive index ($n$), and extinction coefficient ($\kappa$) can be obtained. Figure 3 shows the data acquired with 100-nm-thick, bulk samples on fused quartz. In cases where the material of interest is on a transparent substrate, the process for determining quantities like absorption, reflection, or transmission is substantially easier, further extended the ease with which one can observe excitons and other optical phenomena [51-54]. Excitons are generally seen easily through absorption spectra, which are themselves primarily determined by the imaginary component of the dielectric function ($\varepsilon_2$) and related by $\frac{\varepsilon_2}{A(E)} = \frac{hc}{EL}$, where $E$ is the energy and $L$ is the layer thickness [53].

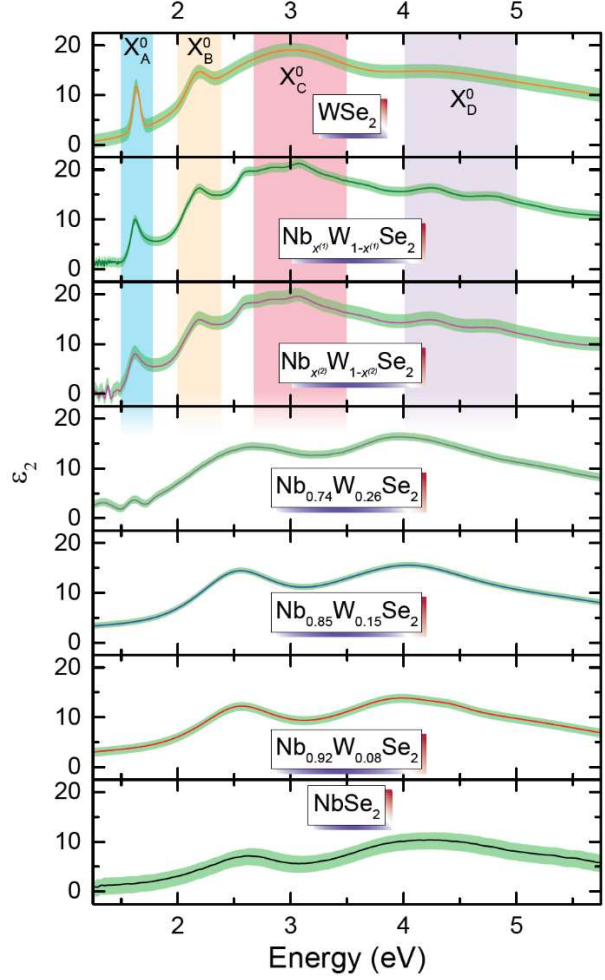

**Fig. 3.** SE measurements of Ψ and Δ are converted such that the imaginary part of the dielectric function $\varepsilon_2(E)$ is shown for each of seven compositions of $Nb_xW_{1-x}Se_2$ alloy as a function of energy. As the Nb concentration $x$ is increased from 0 to 1, $\varepsilon_2$ transitions from semiconducting $WSe_2$ to metallic $NbSe_2$. The phenomena to note are the changes in the *A* and *B* exciton, labelled $X^0_A$ (azure region) and $X^0_B$ (orange region), respectively. The C and D transitions are also labelled as $X^0_C$ (red region) and $X^0_D$ (lavender region), respectively. The shaded green regions around each curve represent a 1σ uncertainty.

Measurements were performed on samples that were freshly cleaved to prevent oxidation effects (see Appendix A). Among the advantages of performing SE on an exfoliated bulk crystal to obtain optical information, two of them are most relevant. First, the thickness does not need to be known to within nanometers because the calculated optical constants (*n* and κ) do not change with perturbations to bulk thickness. Furthermore, the incoming light interacts with a single crystal material as opposed to a layered dielectric. In the latter case, Ψ and Δ would generate a pseudo-dielectric function representing a hybrid material and thus not an accurate representation of the measured sample. In our case, the requirement of precise layer-by-layer analysis is greatly alleviated and Ψ and Δ generate the actual dielectric function of each measured alloy.

Figure 3 shows the results of converting SE measurements of Ψ and Δ to $\varepsilon_2(E)$. The *A* and *B* exciton of WSe$_2$, which are representative of spin-split bands at the K/K' point, are labelled $X^0_A$ (azure region) and $X^0_B$ (orange region) in Fig. 3, respectively. Additionally, the *C* and *D* transitions of WSe$_2$ are respectively labelled $X^0_C$ (red region) and $X^0_D$ (lavender region) in Fig. 3. The bottom four panels show that as NbSe$_2$ becomes W-rich, diverging from its pure form, $\varepsilon_2$ also gradually increases over the whole spectrum until *x* = 0.74 when the changes in the alloy's local band structure become more substantial to accommodate for the increased concentration of W atoms.

**4.2   Optical Constants and Comments on the Low Nb Concentration**

The Nb concentrations for the two lowest values fell below the EDS limit of detection, but due to observed spectral changes, and most prominently from the broadening of $X^0_A$, introduction of Nb atoms to what would have otherwise been a pure WSe2 crystal is a more likely scenario. We compared our result to the same system in another work in which broadening of $X^0_A$ by approximately 50 meV for Nb-doped WSe$_2$ was observed [33]. Our observed broadening of $X^0_A$ was about 38 meV and 63 meV for $x^{(1)}$ and $x^{(2)}$, respectively, as seen in the two central panels of Fig. 4 (a). It is crucial to stress that though the exact concentration cannot be directly measured as prepared, the fact that $x^{(1)} > x^{(2)}$ suggests that $x^{(1)}$ may represent a slightly smaller concentration since the alloy more closely resembles its pure counterpart (WSe$_2$).

To expand on the optical properties of these alloys, more practical representations of how light propagates through these were calculated. In Fig. 4 (b), the full dielectric function was transformed into optical constants and were calculated for each of the seven compositions, with *n* shown in blue (left vertical axis) and κ in pink (right vertical axis). These constants can be easily applied to determine additional quantities such as the absorption, reflection, or transmission spectra.

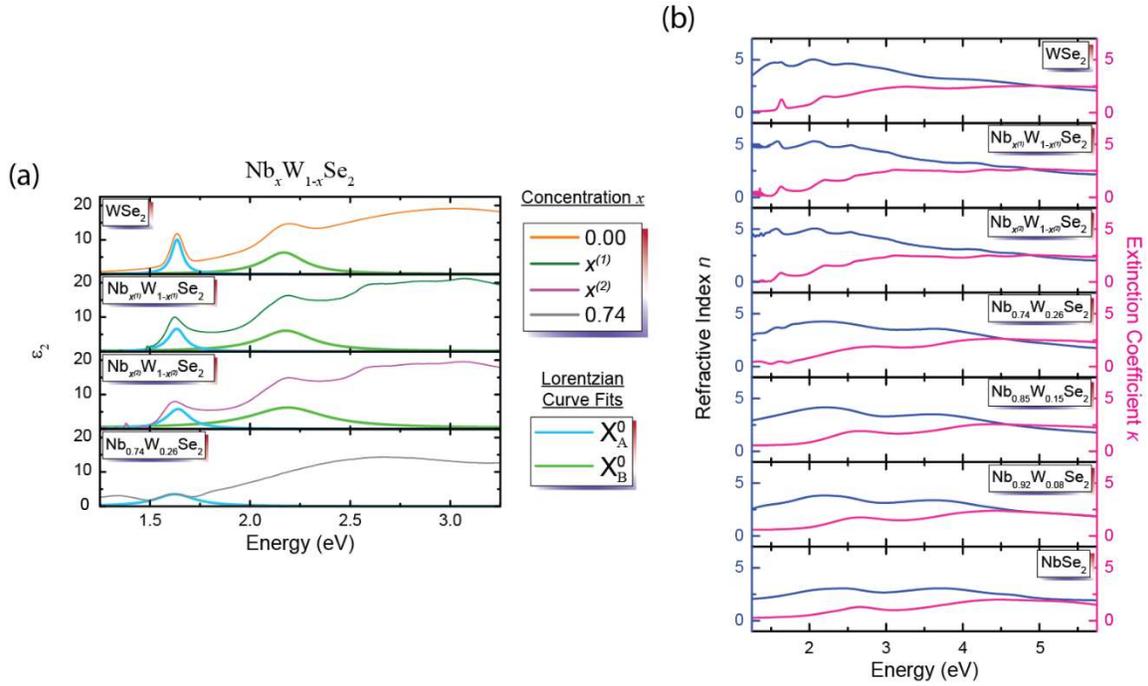

**Fig. 4.** (a) A spectral region covering energies between 1.25 eV and 3.25 eV is shown for four of seven compositions. Lorentzian fits are applied to each curve to extract the position and width of the *A* and *B* excitons, labeled $X^0_A$ (light blue curve) and $X^0_B$ (light green curve), respectively. The *C* and *D* transitions are not shown for clarity but have been included in the fitting procedure. (b) Optical constants are calculated for each of the seven compositions, with the refractive index *n* shown in blue (left vertical axis) and the extinction coefficient κ in pink (right vertical axis).

## 5. Conclusions

In summary, we performed several growths of alloyed crystals with stoichiometric compositions between pure forms of $NbSe_2$ and $WSe_2$. An optical analysis of those alloys was performed by utilizing energy-dispersive X-ray spectroscopy, Raman spectroscopy, and spectroscopic ellipsometry. The imaginary dielectric function was extracted from the ellipsometric data, revealing the extent of spectral changes present in the alloys as a function of transition metal concentration.

## 6. Appendix A

More details regarding the oxidation effects present in exfoliated flakes from the CVT $NbSe_2$ crystal are provided here. Photoluminescence (PL) spectra are visible for the material after one day of exposure to air, highlighting the importance of encapsulation, fast measurement, or immediate placement of freshly exfoliated flakes into an oxygen-free environment. Figure 5 summarizes the PL results as a function of power for both CVT and commercially-obtained flakes. The PL peak arises from the presence of $NbO_2$ and the rate of oxidation appears to scale non-linearly with power [55]. All measurements use flakes that were significantly less exposed to air than the full day after which oxidation PL becomes noticeable.

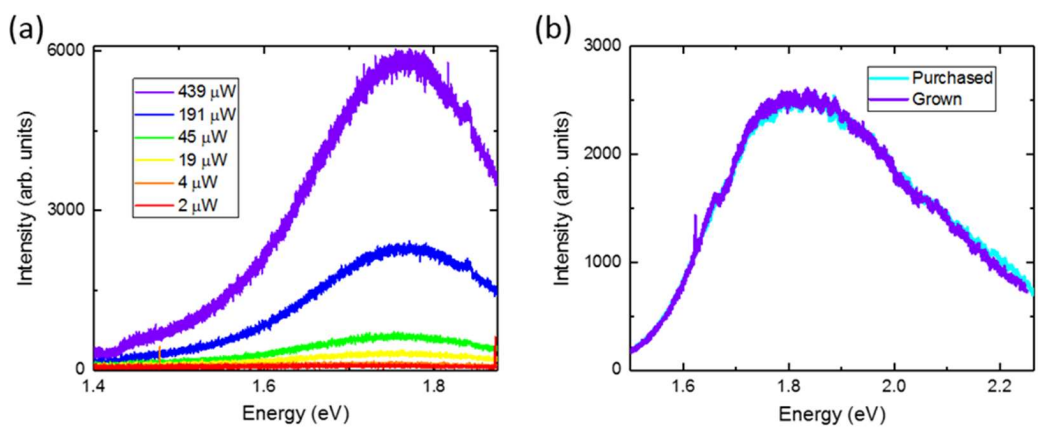

**Fig. 5.** (a) The power-dependent PL spectra are shown here to verify that the large feature around 1.75 eV is likely representative of $NbO_2$ formation resulting from the exposure of the CVT-grown crystal to air for one day. This highlights the importance of performing all optical measurements immediately after exfoliation. (b) When comparing the PL spectrum of the CVT-grown (purple) and commercially-obtained crystal (cyan), both experience the same level of oxidation after one day in air. This supports the idea that the air sensitivity of the crystals is likely inherent to the material.

## Acknowledgments

A.F.R. and H.M.H. have contributed equally to this work. S.K. and A.V.D. provided crystals. The manuscript was written through contributions of all authors. All authors have given approval to the final version of the manuscript. A.F.R and H.M.H. would like to thank the National Research Council's Research Associateship Program for the opportunity. S.K. acknowledges support from the U.S. Department of Commerce, National Institute of Standards and Technology under the financial assistance award 70NANB18H155 and would like to thank Theiss Research for the opportunity. The authors declare no competing interest.

## 7. References

[1] Novoselov KS, Jiang D, Schedin F, Booth TJ, Khotkevich VV, Morozov SV, Geim AK. Two-dimensional atomic crystals. Proceedings of the National Academy of Sciences. 2005 Jul 26;102(30):10451-3. https://doi.org/10.1073/pnas.0502848102

[2] Butler SZ, Hollen SM, Cao L, Cui Y, Gupta JA, Gutiérrez HR, Heinz TF, Hong SS, Huang J, Ismach AF, Johnston-Halperin E. Progress, challenges, and opportunities in two-dimensional materials beyond graphene. ACS nano. 2013 Mar 26;7(4):2898-926. https://doi.org/10.1021/nn400280c

[3] Das S, Robinson JA, Dubey M, Terrones H, Terrones M. Beyond graphene: progress in novel two-dimensional materials and van der Waals solids. Annual Review of Materials Research. 2015 Jul 1;45:1-27. https://doi.org/10.1146/annurev-matsci-070214-021034

[4] Chhowalla M, Shin HS, Eda G, Li LJ, Loh KP, Zhang H. The chemistry of two-dimensional layered transition metal dichalcogenide nanosheets. Nature chemistry. 2013 Apr;5(4):263. https://doi.org/10.1038/nchem.1589

[5] Tonndorf P, Schmidt R, Böttger P, Zhang X, Börner J, Liebig A, Albrecht M, Kloc C, Gordan O, Zahn DR, de Vasconcellos SM. Photoluminescence emission and Raman response of monolayer MoS 2, MoSe 2, and WSe 2. Optics express. 2013 Feb 25;21(4):4908-16. https://doi.org/10.1364/CLEO_QELS.2013.QTu1D.1

[6] Li H, Lu G, Wang Y, Yin Z, Cong C, He Q, Wang L, Ding F, Yu T, Zhang H. Mechanical Exfoliation and Characterization of Single-and Few-Layer Nanosheets of WSe2, TaS2, and TaSe2. Small. 2013 Jun 10;9(11):1974-81. https://doi.org/10.1002/smll.201202919

[7] Mak KF, Lee C, Hone J, Shan J, Heinz TF. Atomically thin MoS 2: a new direct-gap semiconductor. Physical review letters. 2010 Sep 24;105(13):136805. https://doi.org/10.1103/PhysRevLett.105.136805

[8] Berkelbach TC, Hybertsen MS, Reichman DR. Theory of neutral and charged excitons in monolayer transition metal dichalcogenides. Physical Review B. 2013 Jul 25;88(4):045318. https://doi.org/10.1103/PhysRevB.88.045318

[9] Moody G, Dass CK, Hao K, Chen CH, Li LJ, Singh A, Tran K, Clark G, Xu X, Berghäuser G, Malic E. Intrinsic homogeneous linewidth and broadening mechanisms of excitons in monolayer transition metal dichalcogenides. Nature communications. 2015 Sep 18;6:8315. https://doi.org/10.1038/ncomms9315

[10] Qiu DY, Felipe H, Louie SG. Optical spectrum of MoS 2: many-body effects and diversity of exciton states. Physical review letters. 2013 Nov 20;111(21):216805. https://doi.org/10.1103/PhysRevLett.111.216805

[11] Hsu YT, Vaezi A, Fischer MH, Kim EA. Topological superconductivity in monolayer transition metal dichalcogenides. Nature communications. 2017 Apr 11;8:14985. https://doi.org/10.1038/ncomms14985

[12] Manzeli S, Ovchinnikov D, Pasquier D, Yazyev OV, Kis A. 2D transition metal dichalcogenides. Nature Reviews Materials. 2017 Aug;2(8):17033. https://doi.org/10.1038/natrevmats.2017.33

[13] Wang G, Chernikov A, Glazov MM, Heinz TF, Marie X, Amand T, Urbaszek B. Colloquium: Excitons in atomically thin transition metal dichalcogenides. Reviews of Modern Physics. 2018 Apr 4;90(2):021001. https://doi.org/10.1103/RevModPhys.90.021001

[14] Wang QH, Kalantar-Zadeh K, Kis A, Coleman JN, Strano MS. Electronics and optoelectronics of two-dimensional transition metal dichalcogenides. Nature nanotechnology. 2012 Nov;7(11):699. https://doi.org/10.1038/nnano.2012.193

[15] Wilson JA, Yoffe AD. The transition metal dichalcogenides discussion and interpretation of the observed optical, electrical and structural properties. Advances in Physics. 1969 May 1;18(73):193-335. https://doi.org/10.1080/00018736900101307

[16] Neville RA, Evans BL. The Band Edge Excitons in 2H☐ MoS2. physica status solidi (b). 1976 Feb 1;73(2):597-606. https://doi.org/10.1002/pssb.2220730227

[17] Mak KF, Shan J, Heinz TF. Seeing many-body effects in single-and few-layer graphene: observation of two-dimensional saddle-point excitons. Physical review letters. 2011 Jan 25;106(4):046401. https://doi.org/10.1103/PhysRevLett.106.046401

[18] Splendiani A, Sun L, Zhang Y, Li T, Kim J, Chim CY, Galli G, Wang F. Emerging photoluminescence in monolayer MoS2. Nano letters. 2010 Mar 15;10(4):1271-5. https://doi.org/10.1021/nl903868w

[19] Chatterjee U, Zhao J, Iavarone M, Di Capua R, Castellan JP, Karapetrov G, Malliakas CD, Kanatzidis MG, Claus H, Ruff JP, Weber F. Emergence of coherence in the charge-density wave state of 2H-NbSe 2. Nature communications. 2015 Feb 17;6:6313. https://doi.org/10.1038/ncomms7313

[20] Zhu X, Guo Y, Cheng H, Dai J, An X, Zhao J, Tian K, Wei S, Zeng XC, Wu C, Xie Y. Signature of coexistence of superconductivity and ferromagnetism in two-dimensional NbSe 2 triggered by surface molecular adsorption. Nature communications. 2016 Apr 4;7:11210. https://doi.org/10.1038/ncomms11210

[21] Xi X, Zhao L, Wang Z, Berger H, Forró L, Shan J, Mak KF. Strongly enhanced charge-density-wave order in monolayer NbSe 2. Nature nanotechnology. 2015 Sep;10(9):765. https://doi.org/10.1038/nnano.2015.143

[22] Hill HM, Chowdhury S, Simpson JR, Rigosi AF, Newell DB, Berger H, Tavazza F, Walker AR. Phonon origin and lattice evolution in charge density wave states. Physical Review B. 2019 May 20;99(17):174110.

[23] Suderow H, Tissen VG, Brison JP, Martínez JL, Vieira S. Pressure Induced Effects on the Fermi Surface of Superconducting 2 H− NbSe 2. Physical review letters. 2005 Sep 9;95(11):117006. https://doi.org/10.1103/PhysRevLett.95.117006

[24] Arguello CJ, Chockalingam SP, Rosenthal EP, Zhao L, Gutiérrez C, Kang JH, Chung WC, Fernandes RM, Jia S, Millis AJ, Cava RJ. Visualizing the charge density wave transition in 2 H-NbSe 2 in real space. Physical Review B. 2014 Jun 11;89(23):235115. https://doi.org/10.1103/PhysRevB.89.235115

[25] El-Bana MS, Wolverson D, Russo S, Balakrishnan G, Paul DM, Bending SJ. Superconductivity in two-dimensional NbSe2 field effect transistors. Superconductor Science and Technology. 2013 Nov 12;26(12):125020. https://doi.org/10.1088/0953-2048/26/12/125020

[26] Xi X, Berger H, Forró L, Shan J, Mak KF. Gate Tuning of Electronic Phase Transitions in Two-Dimensional NbSe 2. Physical review letters. 2016 Aug 29;117(10):106801. https://doi.org/10.1103/PhysRevLett.117.106801

[27] Lian CS, Si C, Duan W. Unveiling charge-density wave, superconductivity, and their competitive nature in two-dimensional NbSe2. Nano letters. 2018 Apr 13;18(5):2924-9. https://doi.org/10.1021/acs.nanolett.8b00237

[28] Zhou Y, Wang Z, Yang P, Zu X, Yang L, Sun X, Gao F. Tensile strain switched ferromagnetism in layered NbS2 and NbSe2. Acs Nano. 2012 Oct 22;6(11):9727-36. https://doi.org/10.1021/nn303198w

[29] Flicker F, Van Wezel J. Charge order from orbital-dependent coupling evidenced by NbSe 2. Nature communications. 2015 May 7;6:7034. https://doi.org/10.1038/ncomms8034

[30] Xi X, Wang Z, Zhao W, Park JH, Law KT, Berger H, Forró L, Shan J, Mak KF. Ising pairing in superconducting NbSe 2 atomic layers. Nature Physics. 2016 Feb;12(2):139. https://doi.org/10.1038/nphys3538

[31] Bawden L, Cooil SP, Mazzola F, Riley JM, Collins-McIntyre LJ, Sunko V, Hunvik KW, Leandersson M, Polley CM, Balasubramanian T, Kim TK. Spin–valley locking in the normal state of a transition-metal dichalcogenide superconductor. Nature communications. 2016 May 23;7:11711. https://doi.org/10.1038/ncomms11711

[32] Gao J, Kim YD, Liang L, Idrobo JC, Chow P, Tan J, Li B, Li L, Sumpter BG, Lu TM, Meunier V. Transition-Metal Substitution Doping in Synthetic Atomically Thin Semiconductors. Advanced Materials. 2016 Nov;28(44):9735-43. https://doi.org/10.1002/adma.201601104

[33] Hsu HP, Lin DY, Jheng JJ, Lin PC, Ko TS. High Optical Response of Niobium-Doped WSe2-Layered Crystals. Materials. 2019 Jan;12(7):1161. https://doi.org/10.3390/ma12071161

[34] Cho B, Kim AR, Kim DJ, Chung HS, Choi SY, Kwon JD, Park SW, Kim Y, Lee BH, Lee KH, Kim DH. Two-dimensional atomic-layered alloy junctions for high-performance wearable chemical sensor. ACS applied materials & interfaces. 2016 Jul 19;8(30):19635-42. https://doi.org/10.1021/acsami.6b05943

[35] Kim AR, Kim Y, Nam J, Chung HS, Kim DJ, Kwon JD, Park SW, Park J, Choi SY, Lee BH, Park JH. Alloyed 2D metal–semiconductor atomic layer junctions. Nano letters. 2016 Feb 10;16(3):1890-5. https://doi.org/10.1021/acs.nanolett.5b05036

[36] Kim Y, Kim AR, Yang JH, Chang KE, Kwon JD, Choi SY, Park J, Lee KE, Kim DH, Choi SM, Lee KH. Alloyed 2D metal–semiconductor heterojunctions: origin of interface states reduction and schottky barrier lowering. Nano letters. 2016 Aug 25;16(9):5928-33. https://doi.org/10.1021/acs.nanolett.6b02893

[37] Méasson MA, Gallais Y, Cazayous M, Clair B, Rodiere P, Cario L, Sacuto A. Amplitude Higgs mode in the 2H−NbSe2 superconductor. Physical Review B. 2014 Feb 18;89(6):060503. https://doi.org/10.1103/PhysRevB.89.060503

[38] Mialitsin A. Fano line shape and anti-crossing of Raman active E2g peaks in the charge density wave state of NbSe2. Journal of Physics and Chemistry of Solids. 2011 May 1;72(5):568-71. https://doi.org/10.1016/j.jpcs.2010.10.044

[39] Dordevic SV, Basov DN, Dynes RC, Ruzicka B, Vescoli V, Degiorgi L, Berger H, Gaál R, Forró L, Bucher E. Optical properties of the quasi-two-dimensional dichalcogenides 2H-TaSe and 2H-NbSe. The European Physical Journal B-Condensed Matter and Complex Systems. 2003 May 1;33(1):15-23. https://doi.org/10.1140/epjb/e2003-00136-1

[40] Del Corro E, Terrones H, Elias A, Fantini C, Feng S, Nguyen MA, Mallouk TE, Terrones M, Pimenta MA. Excited excitonic states in 1L, 2L, 3L, and bulk WSe2 observed by resonant Raman spectroscopy. ACS nano. 2014 Sep 4;8(9):9629-35. https://doi.org/10.1021/nn504088g

[41] Hill HM, Rigosi AF, Krylyuk S, Tian J, Nguyen NV, Davydov AV, Newell DB, Walker AR. Comprehensive optical characterization of atomically thin NbSe2. Physical Review B. 2018 Oct 5;98(16):165109. https://doi.org/10.1103/PhysRevB.98.165109

[42] Zeng H, Liu GB, Dai J, Yan Y, Zhu B, He R, Xie L, Xu S, Chen X, Yao W, Cui X. Optical signature of symmetry variations and spin-valley coupling in atomically thin tungsten dichalcogenides. Scientific reports. 2013 Apr 11;3:1608. https://doi.org/10.1038/srep01608

[43] Cui Q, Ceballos F, Kumar N, Zhao H. Transient absorption microscopy of monolayer and bulk WSe2. ACS nano. 2014 Feb 24;8(3):2970-6. https://doi.org/10.1021/nn500277y

[44] Guan J, Chuang HJ, Zhou Z, Tománek D. Optimizing charge injection across transition metal dichalcogenide heterojunctions: Theory and experiment. ACS nano. 2017 Mar 23;11(4):3904-10. https://doi.org/10.1021/acsnano.7b00285

[45] Bougouma M, Guel B, Segato T, Legma JB, Ogletree MP. The structure of niobium-doped MoSe2 and WSe2. Bulletin of the Chemical Society of Ethiopia. 2008;22(2). https://doi.org/10.4314/bcse.v22i2.61289

[46] Pandey SK, Alsalman H, Azadani JG, Izquierdo N, Low T, Campbell SA. Controlled p-type substitutional doping in large-area monolayer WSe2 crystals grown by chemical vapor deposition. Nanoscale. 2018 Nov 22;10(45):21374-85. https://doi.org/10.1039/C8NR07070A

[47] Lin DY, Jheng JJ, Ko TS, Hsu HP, Lin CF. Doping with Nb enhances the photoresponsivity of WSe2 thin sheets. AIP Advances. 2018 May 14;8(5):055011. https://doi.org/10.1063/1.5024570

[48] Kruskopf M, Rigosi AF, Panna AR, Marzano M, Patel D, Jin H, Newell DB, Elmquist RE. Next-generation crossover-free quantum Hall arrays with superconducting interconnections. Metrologia. 2019 Oct 10;56(6):065002.

[49] Rigosi AF, Liu CI, Wu BY, Lee HY, Kruskopf M, Yang Y, Hill HM, Hu J, Bittle EG, Obrzut J, Walker AR. Examining epitaxial graphene surface conductivity and quantum Hall device stability with Parylene passivation. Microelectronic engineering. 2018 Jul 5;194:51-5.

[50] Rigosi AF, Patel D, Marzano M, Kruskopf M, Hill HM, Jin H, Hu J, Walker AR, Ortolano M, Callegaro L, Liang CT. Atypical quantized resistances in millimeter-scale epitaxial graphene pn junctions. Carbon. 2019 Dec 1;154:230-7.

[51] Zhao W, Ghorannevis Z, Amara KK, Pang JR, Toh M, Zhang X, Kloc C, Tan PH, Eda G. Lattice dynamics in mono-and few-layer sheets of WS2 and WSe2. Nanoscale. 2013;5(20):9677-83. https://doi.org/10.1039/C3NR03052K

[52] Hill HM, Rigosi AF, Chowdhury S, Yang Y, Nguyen NV, Tavazza F, Elmquist RE, Newell DB, Walker AR. Probing the dielectric response of the interfacial buffer layer in epitaxial graphene via optical spectroscopy. Physical Review B. 2017 Nov 28;96(19):195437. https://doi.org/10.1103/PhysRevB.96.195437

[53] Keçik D, Bacaksiz C, Senger RT, Durgun E. Layer-and strain-dependent optoelectronic properties of hexagonal AlN. Physical Review B. 2015 Oct 9;92(16):165408. https://doi.org/10.1103/PhysRevB.92.165408

[54] Rigosi AF, Hill HM, Glavin NR, Pookpanratana SJ, Yang Y, Boosalis AG, Hu J, Rice A, Allerman AA, Nguyen NV, Hacker CA. Measuring the dielectric and optical response of millimeter-scale amorphous and hexagonal boron nitride films grown on epitaxial graphene. 2D Materials. 2017 Dec 13;5(1):011011. https://doi.org/10.1088/2053-1583/aa9ea3

[55] Zhao Y, Zhang Z, Lin Y. Optical and dielectric properties of a nanostructured NbO2 thin film prepared by thermal oxidation. Journal of Physics D: Applied Physics. 2004 Dec 2;37(24):3392. https://doi.org/10.1088/0022-3727/37/24/006